\documentclass{article}
\usepackage{color}
\usepackage{ifthen}
\usepackage{calc}

\usepackage{morefloats}
\usepackage[inline]{trackchanges}
\addeditor{ref1}
\usepackage[utf8]{inputenc}
\usepackage{authblk}
\usepackage[dvips]{graphicx}
\graphicspath{{noiseimages/}}
\usepackage{xcolor}
\usepackage[T2A]{fontenc}
\usepackage{tikz}
\usepackage{pgfplots}
\usetikzlibrary{intersections, pgfplots.fillbetween}
\usetikzlibrary{decorations.pathmorphing}
\usepackage{amsmath}
\usepackage{amssymb}
\usepackage{float}
\usepackage{physics}
\usepackage{mathtext}
\usepackage{a4wide}
\usepackage{hyperref}
\usepackage{lineno}
\definecolor{urlcolor}{HTML}{990000}
\definecolor{linkcolor}{HTML}{005F5F}
\definecolor{urlcolor}{HTML}{FF0000}
\definecolor{linkcolor}{HTML}{FF0000}
\hypersetup{pdfstartview=FitH,  linkcolor=linkcolor,urlcolor=urlcolor, colorlinks=true,citecolor=blue}

\author[1,2]{E.T.Akhmedov}
\author[1,2]{K.Kazarnovskii}
\affil[1]{Moscow Institute of Physics and Technology, Institutskii per. 9, 141700, Dolgoprudny, Russia}
\affil[2]{ Institute for Theoretical and Experimental Physics, B. Cheremushkinskaya 25, 117218, Moscow, Russia}
\title{\textcolor{black}{Thermalization with non--zero initial anomalous quantum averages}}

\begin{document}


\maketitle

\begin{abstract}
We discuss the thermalization process in the kinetic approximation in the presence of non--zero initial anomalous quantum expectation values on top of an initial non--planckian (non--thermal) level population. Namely we derive a system of ``kinetic'' equations for the level population and anomalous expectation values in four--dimensional massive scalar field theory with $\varphi^4$ self--interaction. We show analytically in the linear approximation that for their small initial values the anomalous quantum averages relax down to zero. Furthermore, we show analytically that this system does not have an equilibrium solution with non--zero time independent anomalous expectation value.
\end{abstract}

{\bf 1.} Understanding of quantum field theory dynamics over strong field backgrounds demands consideration of correlation functions over quantum states with non--zero anomalous quantum expectation values or anomalous averages. We use here the standard terminology from BCS theory of superconductors \cite{AGD}, \cite{LL9}, where anomalous averages play the key role: we name as the anomalous averages the expectation values of the form $Tr[\overset{\wedge}{\rho}a_{\Vec{q}} \, a_{\Vec{q}'}]\equiv \expval{a_{q}a_{q'}}$, where $\hat{\rho}$ is the density matrix characterizing the state of the theory and $a_{\Vec{q}}$ are the annihilation operators in the theory; the trace is taken over the Hilbert space of the theory. For the standard Fock space states (e.g. Poincare invariant ground state) the anomalous averages are always zero. However, in quantum field theory over strong background fields even for zero initial values the anomalous averages are generated dynamically in loop corrections. This happens e.g. in expanding universe \cite{Krotov:2010ma}, \cite{Moreau:2018lmz}, \cite{Guilleux:2015pma}, \cite{Akhmedov:2013vka}, \cite{Akhmedov:2019cfd}, during collapse process \cite{Akhmedov:2015xwa}, in the presence of strong electric fields \cite{Akhmedov:2014hfa}, \cite{Akhmedov:2014doa} and in the presence of background scalar fields \cite{Akhmedov:2020haq}. See \cite{Akhmedov:2021rhq} for a general discussion. 

At the same time, quantum field dynamics in flat space--time with non--trivial anomalous expectation values is poorly studied. That is mainly because to obtain non-zero anomalous averages one has to consider some sort of coherent states. Meanwhile consideration of states with non--zero anomalous averages in flat space quantum field theory can be a suitable playground for the development of methods to work in similar situations over strong background fields. Concretely in this paper we consider the thermalization process in the massive scalar field theory with $\varphi^4$ self--interaction in four--dimensional Minkowski space--time.

The thermalization process in the kinetic approximation with zero anomalous expectation values was considered in many places (see e.g. textbooks \cite{LL10}, \cite{Zubarev}, \cite{Mattuck}). However, to the best of our knowledge, the thermalization process in the kinetic approximation with non--zero anomalous averages was not considered.

In this paper we consider an initial state, containing non--planckian (non--thermal) level population, which is expressed via $Tr[\overset{\wedge}{\rho}a^+_{\Vec{q}}a_{\Vec{q}'}] \equiv \expval{a^+_{q}a_{q'}}$, and anomalous averages, $Tr[\overset{\wedge}{\rho}a_{\Vec{q}}a_{\Vec{q}'}]\equiv \expval{a_{q}a_{q'}}$.
Our goal is to show that such a state will evolve in time towards the planckian level population, $n_p=\frac{1}{e^{\beta\epsilon_p}-1}$, and zero anomalous average. Namely the final state of the thermalization process is described by the planckian distribution of modes, $n_p$, over the Poincare invariant ground state. 

To show this phenomenon we derive a system of kinetic equations for the level population and anomalous averages. Then we show that this system has a solution with zero anomalous averages and planckian level population only for such modes, which diagonalize the free Hamiltonian. To derive the system of generalized kinetic equations we use the standard text book methods, which can be found e.g. in \cite{Zubarev} and \cite{Mattuck}. To have the analytic headway in finding solutions of this system, we consider spatially homogeneous states and small initial anomalous averages. But we also derive the system of kinetic equations for $n_p$ and $\chi_p$ without the latter assumptions.

{\bf 2.} In this note we consider the four--dimensional real massive self--interacting scalar field theory in flat space--time:

\begin{align}
\mathcal{L}=\frac{1}{2}(\partial_{\mu}\varphi)^2-\frac{m^2}{2}\varphi^2-\frac{\lambda}{4!}\varphi^4.\label{lagr}
\end{align}
The signature of the metric is $(+---)$. (We work in the units $\hbar = 1 = c$.)
We are interested only in the infrared (IR) effects and assume that all coupling constants in this Lagrangian acquire their physical (UV renormalized) values.

It is well known that a gas of quanta, whose dynamics is described by the Lagrangian (\ref{lagr}), will thermalize for any physically reasonable initial level population $Tr[\overset{\wedge}{\rho}a^+_{\Vec{q}}a_{\Vec{q}'}] \equiv \expval{a^+_{q}a_{q'}} \sim n^0_q \, \delta(\vec{q}-\vec{q}')$. Below we will review how to make this observation theoretically for the situation which is close to the equilibrium. Namely we will show that in time the distribution will become planckian $n_q = \frac{1}{e^{\beta\epsilon_q}-1}$, with the temperature $1/\beta$ following from the initial data and $\epsilon_q = \sqrt{\vec{q}^2 + m^2}$ being the energy of the mode with the momentum $\vec{q}$.

The goal of this paper is to see the thermalization with a non--planckian initial level population, when there is also non--zero initial anomalous quantum expectation values, $Tr[\overset{\wedge}{\rho}a_{\Vec{q}}a_{\Vec{q}'}]\equiv \expval{a_{q}a_{q'}} \sim \chi^0_q \, \delta(\vec{q}+\vec{q}')$. In this paper we make analytic observations for small initial values of the anomalous averages, i.e. also when the state of the theory is close to equilibrium.

There are two ways to introduce non--zero initial values of the anomalous averages, which are related to each other via canonical transformations. The first option is to use the standard modes:

\begin{align}\label{decomp}
    \varphi(x,t)=\int \frac{d^3p}{(2\pi)^3}\left[a_p\frac{e^{i(\Vec{p}\Vec{x}-\epsilon_pt)}}{\sqrt{2\epsilon_p}} + a^+_p\frac{e^{-i(\Vec{p}\Vec{x}-\epsilon_pt)}}{\sqrt{2\epsilon_p}}\right],\quad \epsilon_p=\sqrt{\Vec{p\ }^2+m^2},
\end{align}
i.e. from the very beginning one starts with the modes, which diagonalize the free Hamiltonian following from the Lagrangian (\ref{lagr}). Then, one considers such an initial state in which:

\begin{align}\label{initial}
\expval{a^+_{q}a_{q'}} = n^0_q \, \delta\left(\vec{q} - \vec{q}'\right)\quad {\rm and} \quad \expval{a_{q}a_{q'}} = \chi^0_q \, \delta\left(\vec{q} + \vec{q}'\right),
\end{align}
where both $n^0_q$ and $\chi^0_q$ are not zero and $n^0_q$ is not equal to the planckian distribution. For simplicity we begin with the consideration of spatially homogeneous states. Inhomogeneous situations will be briefly discussed below. Also in this paper we consider only such states for which Wick's theorem does work. Furthermore, to make analytical considerations we have to consider small initial $\chi^0_q$.

The second option is to consider the decomposition of the field operator as follows

\begin{align}\label{wrong1}
    \varphi(x,t)=\int \frac{d^3p}{(2\pi)^3}\left[b_p \, \frac{f_p(t,x)}{\sqrt{2\epsilon_p}} + b^+_p \, \frac{f^{*}_p(t,x)}{\sqrt{2\epsilon_p}}\right],
\end{align}
where the modes have the form:

\begin{align}\label{wrong2}
    f_p(t,x)=u_p e^{i(\Vec{p}\Vec{x}-\epsilon_pt)}+v_pe^{-i(\Vec{p}\Vec{x}-\epsilon_pt)}.
\end{align}
To obtain the canonical commutation relations for $b_p$ with $b^+_p$, and for the field operator $\varphi(x,t)$ with its conjugate momentum, constants $u_p$ and $v_p$ should obey the relation $|u_p|^2 - |v_p|^2 = 1$.

Furthermore, to obtain the proper UV (Hadamard) behaviour of the correlation functions it is necessary to demand the condition $\lim_{|\Vec{p}|\to\infty} v_p =0$. In fact, then for very large momenta the modes $f_p(t,x)$ behave in the standard way, i.e. as those in (\ref{decomp}). As a result, correlation functions have the standard UV singularities. That is necessary for the proper UV renormalization of physical quantities in the loops. E.g. only in such a case the beta--function of the coupling constant $\lambda$ will depend only on the kinematic number of the degrees of freedom rather than on the values of $u_p$ and $v_p$.

For such a mode decomposition as (\ref{wrong1}) and (\ref{wrong2}) one can consider e.g. an initial state of the form:

\begin{align}\label{blevdis}
    Tr[\overset{\wedge}{\rho}b^+_qb_p]=\delta^3(\Vec{p}-\Vec{q}) \, n^0_p,\quad Tr[\overset{\wedge}{\rho}b_pb_q]=0.
\end{align}
But performing the Bogoliubov transformation to the standard modes ($\ref{decomp}$):

\begin{align}\label{Bogoluib}
    a_p=u_pb_p+v_pb^+_p,\quad a^+_p=u_pb^+_p+v_pb_p,
\end{align}
we can reduce the problem to the one with the mode decomposition (\ref{decomp}) and a state of the form (\ref{initial}), where the initial level population and anomalous averages are equal to:

\begin{align}
\bar{n}^0_q = n^0_q + O(v_q^2),\quad \bar{\chi}^0_q=u_pv_p\left(1 + 2 \, n^0_q\right).
\end{align}
Again for simplicity we have assumed that $v_q$ is small for all $q$.

Of course (\ref{wrong2}) is not the most generic form of the modes, because Bogoliubov rotations are not the most generic canonical transformations. But for the consideration of such states which are very close to the spatial homogeneity it is sufficient to consider the modes of the form (\ref{wrong2}).

{\bf 3.}  Thus, we restrict our attention to the first way, (\ref{decomp}) and (\ref{initial}), of setting up the initial anomalous expectation values. In the theory (\ref{lagr}) the free Hamiltonian has the standard form. Meanwhile the normal ordered interaction term is as follows:

\begin{eqnarray}
    H_{int}(t)=\frac{\lambda}{4!}\int \frac{d^3p_1 \dots d^3p_4}{(2\pi)^{12}}\frac{1}{4\sqrt{\epsilon_{p_1}\epsilon_{p_2}\epsilon_{p_3}\epsilon_{p_4}}} \times \nonumber \\ \times  \delta^3(\Vec{p_1}+\Vec{p_2}-\Vec{p_3}-\Vec{p_4}) \, \Big[a_{p_1}a_{p_2}a_{p_3}a_{p_4}e^{-i(\epsilon_{p_1}+\epsilon_{p_2}+\epsilon_{p_3}+\epsilon_{p_4}) \, t } + \nonumber \\ + 4 \, a^+_{p_1}a_{p_2}a_{p_3}a_{p_4} \, e^{i(\epsilon_{p_1}-\epsilon_{p_2}-\epsilon_{p_3}-\epsilon_{p_4}) \, t} + 6 \, a^+_{p_1}a^+_{p_2}a_{p_3}a_{p_4} \, e^{i(\epsilon_{p_1}+\epsilon_{p_2}-\epsilon_{p_3}-\epsilon_{p_4}) \, t } + \nonumber \\ + 4\, a^+_{p_1}a^+_{p_2}a^+_{p_3}a_{p_4} \, e^{i(\epsilon_{p_1}+\epsilon_{p_2}+\epsilon_{p_3}-\epsilon_{p_4})\, t} + a^+_{p_1}a^+_{p_2}a^+_{p_3}a^+_{p_4} \, e^{i(\epsilon_{p_1}+\epsilon_{p_2}+\epsilon_{p_3}+\epsilon_{p_4})\, t}\Big].\label{Hint}
\end{eqnarray}
Note that such an interaction Hamiltonian does not commute with the total particle number unlike the one in superconductor BCS theory.

To understand the thermalization process one writes the interaction picture evolution equations for the following expectation values:

\begin{eqnarray}
\frac{d}{dt} \expval{a^+_{q}a_{q'}} = i \, \Big\langle \left[H_{int}(t), \, a^+_{q}a_{q'}\right]\Big\rangle, \quad
\frac{d}{dt} \expval{a_{q}a_{q'}} = i \, \Big\langle \left[H_{int}(t), \, a_{q}a_{q'}\right]\Big\rangle.\label{begin}
\end{eqnarray}
The novel moment that is considered in this paper in comparison with the standard situation is that now we add the equation for $\expval{a_{q}a_{q'}}$ on top of the standard equation for $\expval{a^+_{q}a_{q'}}$. These quantities are related to the level--population and anomalous averages as $\expval{a^+_{q}a_{q'}} = n_q \, \delta\left(\vec{q} - \vec{q}'\right)$ and $\expval{a_{q}a_{q'}} = \chi_q \, \delta\left(\vec{q} + \vec{q}'\right)$, if the average is taken over the spatially homogeneous state. Throughout this paper we always assume that $n_q$ and $\chi_q$ are functions of the modulus of $\vec{q}$.

Consider first the linear order in $\lambda$. At this order, using the Wick's theorem, one obtains\footnote{Throughout this paper we consider only such states, with respect to which the expectation values are taken, for which the Wick's theorem is fulfilled  \cite{Radovskaya:2020lns}. Also at the beginning we consider only such states which respect the spatial translational invariance. Extension to non--invariant states will be considered below.}:

\begin{align}
    \frac{dn_q}{dt} = \frac{i\lambda}{2}\int \frac{d^3p}{4(2\pi)^3 \epsilon_q \epsilon_p} \Bigg[\Big(\chi_q \chi_pe^{- 2 i (\epsilon_q + \epsilon_p) t} - \chi^{*}_q \chi^{*}_p e^{2 i (\epsilon_q + \epsilon_p) t}\Big)+\nonumber \\
    (1+2n_p)\Big(\chi_q e^{-2i\epsilon_q t} - \chi^{*}_q e^{2i\epsilon_q t}\Big) + \Big(\chi^{*}_p \chi_q e^{2i(\epsilon_q-\epsilon_q) t} - \chi_p\chi^{*}_q e^{2i(\epsilon_q-\epsilon_p) t}\Big) \Bigg],\label{lambda1}
\end{align}
and

\begin{align}
    \frac{d\chi_q}{dt} = \frac{-i\lambda}{2}\int\frac{d^3p}{4(2\pi)^3\epsilon_q\epsilon_p} \, \Bigg[2\chi_q(1+2n_p)+(1+2n_q)\chi^{*}_pe^{2i(\epsilon_q+\epsilon_p)t}+(1+2n_q)\chi_pe^{2i(\epsilon_q-\epsilon_p)t} \nonumber \\
  +(1+2n_q) (1+2n_p) e^{2 i \epsilon_q t} + 2 \chi_q \chi^{*}_p e^{2 i \epsilon_p t} + 2 \chi_q \chi_p e^{-2 i \epsilon_p t}\Bigg]. \label{lambda2}
\end{align}
In deriving these equations we have used that $\expval{a^+_{q}a_{q'}} = n_q \, \delta\left(\vec{q} - \vec{q}'\right)$ and $\expval{a_{q}a_{q'}} = \chi_q \, \delta\left(\vec{q} + \vec{q}'\right)$ for translationally invariant states. Here $n_p$ and $\chi_p$ are the level population and anomalous averages at the moment of time $t$, while $n_q^0$ and $\chi_q^0$, which were introduced above, are their values at the initial moment of time. Also we assume the kinetic approximation here, i.e. that $n_p$ and $\chi_q$ are slow functions of time $t$ in comparison with the modes.

Oscillating terms on the right hand sides of (\ref{lambda1}) and (\ref{lambda2}) can be neglected, as they do not lead to the secular growth of $n_q$ and $\chi_q$ after the integration of both sides of these equations over $t$. It means that in the limit $\lambda \to 0$ the change of $n_q$ and $\chi_q$ in time , which is caused by such terms, is negligible. Then, neglecting the oscillating terms, one obtains:
\begin{align}
    \frac{dn_q}{dt} =0
\end{align}
and
\begin{align}
    \epsilon_q \, \frac{d\chi_q}{dt} = \chi_q \,  \frac{-i\lambda}{2} \, \int\frac{d^3p}{4(2\pi)^3\epsilon_p} \, \Bigg[2 \, (1 + 2 \, n_p)\Bigg]. \label{lambda3}
\end{align}
We can get the analytic headway assuming that the system is very close to the equilibrium, i.e. that:

\begin{align}\label{assump}
    \chi^0_q(t=0) \ll 1,\quad n^0_q(t=0) =\frac{1}{e^{\beta\epsilon_q}-1}+\Tilde{n}_q,\quad \Tilde{n}_q\ll \frac{1}{e^{\beta\epsilon_q}-1}.
\end{align}
The first condition here is necessary to linearize the equation for $\chi_q$, if they were nonlinear, as will be the case below in the $\lambda^2$ order.

If one substitutes planckian distribution for $n_p$ into the right hand side (RHS) of eq. (\ref{lambda3}) it is not hard to recognize in it the term responsible for the mass renormalization. Namely, the integral on the RHS of (\ref{lambda3}) is divergent and can be absorbed into the renormalization of $\epsilon_q$. In fact, if one substitutes $\chi_q = \chi_q^{ren} \, e^{i \, 2\, \delta\epsilon_q \, t}$ into eq. (\ref{lambda3}), then $\delta\epsilon_q$ can be adjusted such that (\ref{lambda3}) acquires the form $d\chi_q^{ren}/dt = 0$.

Furthermore, the RHS of eq. (\ref{lambda3}) appears due to the bubble diagram of the $\lambda \, \varphi^4$ theory, which is responsible for the self-energy renormalization. The resummation of such diagrams can be related to the renormalization of the self-energy, which should be done in the process of normal ordering of the interaction term (\ref{Hint}) in the Hamiltonian. This is another argument why the RHS of eq. (\ref{lambda3}) can be attributed to the mass renormalization. Thus, from now on we deal with $\chi_q^{ren}$ with the renormalized mass, but denote it as usual by $\chi_q$.

{\bf 4.} Let us continue with the second order in $\lambda$. Essentially here we briefly review the derivation of the kinetic equations \cite{Zubarev}, \cite{Mattuck}, but in the presence of anomalous averages. Again one starts with (\ref{begin}), but does not truncate (with the use of the Wick contractions) the RHS immideatly after the calculation of the commutator. Namely one uses the commutation relations for the annihilation and creation operators to obtain on the RHS of (\ref{begin}) the expectation values of the operators which contain fourth powers of $a_p$ and $a_p^+$ operators. E.g., the equation for $\expval{a^+_{q}a_{q'}}$ contains such terms as:

\begin{align}\label{example}
     \expval{\big[a^+_{p_1}a^+_{p_2}a^+_{p_3}a_{p_4},a^+_{q}a_{q'}\big]}=\delta^3(\Vec{q}-\Vec{p_4})\expval{a^+_{p_1}a^+_{p_2}a^+_{p_3}a_{q'}}-3\delta^3(\Vec{q}'-\Vec{p_1})\expval{a^+_{q}a^+_{p_2}a^+_{p_3}a_{p_4}}.
\end{align}
Such an expression stands under the integral over $d^3p_1 \dots d^3p_4$ on the RHS of (\ref{begin}), which comes from the definition of $H_{int}$ in (\ref{Hint}). This way one obtains the dynamical equations for $\expval{a^+_{q}a_{q'}}$ and $\expval{a_{q}a_{q'}}$, which are expressed via the expectation values of the quartic in $a$ and $a^+$ operators.

The next step is to derive the interaction picture evolution equations for the obtained quartic in $a$ and $a^+$ operators --- such operators which appear e.g. on the RHS of (\ref{example}). Namely

\begin{eqnarray}
\frac{d}{dt'} \expval{a^+_{q}a^+_{p_2}a^+_{p_3}a_{p_4}} = i \, \Big\langle \left[H_{int}(t'), \, a^+_{q}a^+_{p_2}a^+_{p_3}a_{p_4}\right]\Big\rangle. \label{begin1}
\end{eqnarray}
Continuing this procedure, one obtains the so called Bogoliubov hierarchy \cite{Zubarev}, \cite{Mattuck}. To truncate it and to close the system of the equations for the level population and anomalous averages one uses the kinetic approximation and the Wick contractions. In the previous section we did the contraction on the previous step. 

The key difference with respect to the standard procedure \cite{Zubarev}, \cite{Mattuck} is that now we are taking into account that anomalous averages are also not equal to zero. For example, from the first expectation value on the right hand side of ($\ref{example}$) we obtain the following contribution to the RHS of (\ref{begin}):

\begin{align}
    \int \frac{d^3p_1 \dots d^3p_4}{(2\pi)^{12}} \,  \delta^3(\Vec{p_1}+\Vec{p_2}-\Vec{p_3}-\Vec{p_4}) \, \delta^3(\Vec{q}-\Vec{p_4}) \, \expval{[H_{int}(t'), \, a^+_{p_1}a^+_{p_2}a^+_{p_3}a_{q'}]} = \nonumber \\ = \frac{\lambda}{4!} \, \delta^3\left(\Vec{q}-\Vec{q}'\right) \, \int\frac{d^3p_1d^3p_2d^3p_3}{(2\pi)^9} \, \frac{1}{4\sqrt{\epsilon_{1}\epsilon_{2}\epsilon_{3}\epsilon_{q}}} \times \nonumber \\ \times
    \Bigg\{4! \, \Big[n_q-(1+n_q)\Big] \, \chi^{*}_1\chi^{*}_2\chi^{*}_3 \, e^{i(\epsilon_{1}+\epsilon_{2}+\epsilon_{3}+\epsilon_{q}) \, t'} + \nonumber \\ + 18 \, \cdot4 \, \Big[n_q(1+n_2)-(1+n_q)n_2\Big] \, \chi^{*}_1\chi^{*}_3 \, e^{i(\epsilon_q+\epsilon_1+\epsilon_3-\epsilon_2)\, t'} + \nonumber \\ + 12 \cdot 6 \, \Big[(1+n_2)-n_2\Big] \, \chi_q\chi^{*}_1\chi^{*}_3 \, e^{i(\epsilon_1+\epsilon_3-\epsilon_q-\epsilon_2)t'} + \nonumber \\ + 12\cdot6 \, \Big[n_q(1+n_1)(1+n_2)-(1+n_q)n_1 n_2\Big] \, \chi^{*}_3 \, e^{i(\epsilon_3 + \epsilon_q - \epsilon_1 - \epsilon_2) \, t'} + \nonumber \\ + 18\cdot4 \, \Big[(1+n_1)(1+n_2)-n_1 n_2\Big] \, \chi_q\chi^{*}_3 \, e^{i(\epsilon_3-\epsilon_1-\epsilon_2-\epsilon_q)\, t'} + \nonumber \\ + 6\cdot4  \, \Big[n_q(1+n_1)(1+n_2)(1+n_3)-(1+n_q)n_1 n_2 n_3\Big] \, e^{i(\epsilon_q-\epsilon_1-\epsilon_2- \epsilon_3) \, t'} + \nonumber \\ + 4! \, \chi_q \Big[(1+n_1)(1+n_2)(1+n_3) - n_1 n_2 n_3\Big] \, e^{i(\epsilon_{1}+\epsilon_{2}+\epsilon_{3}+ \epsilon_{q}) \, t'}\Bigg\}, \label{5}
\end{align}
where $n_i \equiv n_{p_i}$, $i = 1,2,3$. In deriving the last expression we have used again the Wick contractions and that $\expval{a^+_{q}a_{q'}} = n_q \, \delta\left(\vec{q} - \vec{q}'\right)$ and $\expval{a_{q}a_{q'}} = \chi_q \, \delta\left(\vec{q} + \vec{q}'\right)$. 

On the next step from eq. (\ref{5}) one finds the time evolution of such operators as $a^+_{p_1}a^+_{p_2}a^+_{p_3}a_{q'}$ via  equations of the form:

\begin{eqnarray}
\expval{a^+_{q}a^+_{p_2}a^+_{p_3}a_{p_4}}(t) = i \, \int_{t_0}^t dt' \Big\langle \left[H_{int}(t'), \, a^+_{q}a^+_{p_2}a^+_{p_3}a_{p_4}\right]\Big\rangle, \label{tprime}
\end{eqnarray}
which then can be substituted into (\ref{begin}) and (\ref{example}). Here $t_0$ is the time after which the interaction term is adiabatically turned on. Please keep in mind that we eventually take the limit $t-t_0 \to \infty$.

The final step is to take the integral over $t'$ in such expressions as (\ref{tprime}) in the kinetic approximation. In this approximation all $n$'s and $\chi$'s are assumed to be slow functions of time in comparison with the oscillating modes. That is a natural assumption in the vicinity of the equilibrium. Then such integrals as in (\ref{tprime}) lead to delta--functions establishing energy conservation in the limit $t-t_0 \to \infty$.

If we now exclude all terms which contain oscillating functions of $t$ and such terms for which arguments of the delta--functions, that establish the energy--momentum conservation, cannot be zero, we get the following system of kinetic equations for $n_p$ and $\chi_p$:

\begin{align}\label{n}
    \epsilon_q \, \frac{d}{dt} n_q = \frac{\lambda^2}{16} \, \int\frac{d^3p_1d^3p_2d^3p_3}{(2\pi)^9 \, \epsilon_1\epsilon_2\epsilon_3} \, \delta^4(\underline{q} + \underline{p}_1-\underline{p}_2-\underline{p}_3) \times \nonumber \\ \times \Big[(1+n_q)(1+n_1)n_2n_3-n_qn_1(1+n_2)(1+n_3)\Big],
\end{align}
and:

\begin{align}\label{kk}
    \epsilon_q \, \frac{d}{dt}\chi_q = \frac{\lambda^2}{16}\int\frac{d^3p_1d^3p_2d^3p_3}{(2\pi)^9\, \epsilon_1\epsilon_2\epsilon_3} \, \delta^4(\underline{q}+\underline{p}_1-\underline{p}_2-\underline{p}_3) \times \nonumber \\ \times \Big\{\chi_q \Big[(1+n_1)n_2n_3 - n_1(1+n_2)(1+n_3)\Big] + 2 \, \chi^{*}_1\chi_2\chi_3 \Big[n_q-(1+n_q)\Big]\Big\},
\end{align}
where $\underline{q} = (\epsilon_q, \, \vec{q})$, $\epsilon_q = \sqrt{\vec{q}^2 + m^2}$ and $\underline{p}_i = (\epsilon_i, \, \vec{p}_i)$, $\epsilon_i = \sqrt{\vec{p}_i^2 + m^2}$, $i = 1,2,3$. It is possible to show that these equations resum a subclass of IR dominant Feynman diagrams in the Schwinger--Keldysh diagrammatic technique \cite{Zubarev}, \cite{Mattuck}, \cite{Kamenev}.

We have derived these equations for spatially homogeneous states. The generalization to slightly spatially inhomogeneous situations is straightforward:  one just has to assume that $n_p$ and $\chi_p$ are functions of $t$ and $\vec{x}$ simultaneously and make the following changes of the derivatives on the left hand side (LHS) of the equations (\ref{n}) and (\ref{kk}):

\begin{align}
    \epsilon_q \frac{d}{dt} \to \epsilon_q \, \partial_t + \vec{q} \, \vec{\partial} = q^\mu \, \partial_\mu.
\end{align}
This concludes the derivation of the system of kinetic equations. We will use it to study the thermalization process in the kinetic approximation, i.e. when the system is close to the equilibrium.

Now one can see that eq. (\ref{n}) for $n_q$ does not contain any terms with anomalous averages $\chi_k$. It is just the well known quantum Boltzmann equation \cite{Zubarev}, \cite{Mattuck}. The RHS of (\ref{n}) is the collision integral which describes competing scattering processes of gaining quanta on the level $\epsilon_p$ and loosing them from the same level. The new feature of the system of kinetic equations (\ref{n}) and (\ref{kk}) is that the RHS of the second equation describes similar kinetic scattering processes in which momentum and energy can be absorbed or generated by the background quantum state of the system.

Let us stress here an important moment relating our formal derivations to real physical processes. In fact, obviously spatially inhomogeneous variants of equations (\ref{n}) and (\ref{kk}) only qualitatively describe situations realized in nature. To obtain equations describing the real physical dynamics quantitatively one just has to modify the rates of various kinetic processes in the collision integrals --- on the RHS of (\ref{n}) and (\ref{kk}). These rates can be found empirically.

Now it is not hard to see that planckian distribution $n_q = \frac{1}{e^{\beta\epsilon_q}-1}$, $\beta = const$ solves (\ref{n}): such a distribution annihilates its both sides. Furthermore, from the Boltzmann's equation follows the second law of thermodynamics. Thus, dynamics described by this equation has a direction. Hence, the Boltzmann's equation predicts the thermalization for an initial non--planckian distribution $n^0_q$.

Now we are ready to observe what happens with $\chi_q$ on the way towards the equilibrium. Let us assume again that at the initial moment of time $t=0$ the system was not very far from the equilibrium, i.e. that the conditions (\ref{assump}) are fulfilled. Again only in such a situation we can make analytical observations. Otherwise one has to solve the system of kinetic equations in question numerically. The second and third conditions in (\ref{assump}) allow one to exclude the situation in which the initial level population generates an increase of $\chi_q$ in the linear approximation.

Under the conditions (\ref{assump}) we obtain, that $n_q$ approximately solves the equation ($\ref{n}$), while from (\ref{kk}) we get:

\begin{align} \label{solch}
     \epsilon_q \, \frac{d}{dt}\chi_q \approx \chi_q \,  \frac{\lambda^2}{16}\int\frac{d^3p_1d^3p_2d^3p_3}{(2\pi)^9\, \epsilon_1\epsilon_2\epsilon_3} \, \delta^4(\underline{q}+\underline{p}_1-\underline{p}_2-\underline{p}_3) \, \Big[(1+n_1)n_2n_3 - n_1(1+n_2)(1+n_3)\Big].
\end{align}
Since for planckian distribution $n_p$ the following relation is true:

\begin{align*}
\delta^4(\underline{q}+\underline{p}_1-\underline{p}_2-\underline{p}_3) \Big[(1+n_q)(1+n_1)n_2n_3-n_qn_1(1+n_2)(1+n_3)\Big]=0,
\end{align*}
one obtains that

\begin{multline}
\delta^4(\underline{q}+\underline{p}_1-\underline{p}_2-\underline{p}_3)\Big[(1+n_1)n_2n_3-n_1(1+n_2)(1+n_3)\Big] = \\
=\delta^4(\underline{q}+\underline{p}_1-\underline{p}_2-\underline{p}_3)\frac{-(1+n_1)n_2n_3}{n_q}\textless 0.
\end{multline}
As a result, the solution of (\ref{solch}) has the form $\chi_q(t)=C_q e^{-\Gamma_q t}$, where the constant $C_q$ is given by the initial conditions and $\Gamma_q > 0$ is defined as:

\begin{align}
   \epsilon_q \Gamma_q=\frac{\lambda^2}{16}\int\frac{d^3p_1d^3p_2d^3p_3}{(2\pi)^9\, \epsilon_1\epsilon_2\epsilon_3} \, \delta^4(\underline{q}+\underline{p}_1-\underline{p}_2-\underline{p}_3)\Big[n_1(1+n_2)(1+n_3) - (1+n_1)n_2n_3  \Big],
\end{align}
It is not hard to see that the integral in this expression is convergent and its sign coincides with the sign of the integrand.

{\bf 5.} So far we have considered the approach to such an equilibrium state in which the anomalous average is zero. Let us show now that in the theory (\ref{lagr}) the equilibrium with non--zero anomalous average is impossible.

Consider an equilibrium state with non-zero anomalous average $\expval{a_{\Vec{p}}a_{\Vec{q}}}=\delta^3(\Vec{p}+\Vec{q})\chi_p$. As this state is stationary the Wightman function for it, $W(x_1,t_1,x_2,t_2)=\expval{\varphi(x_1,t_1)\varphi(x_2,t_2)}$, should depend only on the time difference $\tau=t_2-t_1$, as there should be space and time translational symmetry in equilibrium. For this to be true the anomalous average must depend on time in the following way:

\begin{align}\label{chikap}
    \chi_q(t)=e^{2i \, \epsilon_q \, t}\kappa_q, \quad t=\frac{t_1+t_2}{2}.
\end{align}
This can be seen after the substitution of the mode expansion (\ref{decomp}) of the field operator into the expression of the Wightman function, $W(x_1,t_1,x_2,t_2)=\expval{\varphi(x_1,t_1)\varphi(x_2,t_2)}$. In such a case as (\ref{chikap}) the anomalous average $\chi_q(t)$ is not a slow function of time even at the equilibrium, when $\kappa_q$ is time independent. But $n_p$ and $\kappa_q$ enter into the Wightman function in the combination $N_p = n_p + {\rm Re}\, \kappa_p$ for which the kinetic equation has the standard form (\ref{n}).

Let us stress, however, that there are theories in which formation of a condensate is possible. Non--trivial condensate is directly related to the presence of non--zero anomalous average of certain form. One can keep in mind e.g. the BCS theory, in which the condensate of Cooper pairs corresponds to the situation with non--zero anomalous average or the superfluid He4 or Bose-Einstein condensates \cite{AGD}, \cite{LL9}. However, in contrast to (\ref{lagr}) those theories are non-relativistic; in such theories anomalous averages can be distinguished from the level distributions even in stationary situations and interaction terms in those theories respect the total particle number.

{\bf 6.} Thus, we have derived the system of kinetic equations for the level population and anomalous expectation values in four--dimensional massive scalar field theory with $\varphi^4$ self--interaction. Using this system in the linear approximation we have shown analytically that for their small initial values the anomalous quantum averages relax down to zero.

Furthermore, we have shown that this system does not have an equilibrium solution with non--zero time independent anomalous expectation values. It is instructive to see if the last observation has any relation to the fact that the Coleman--Weinberg mechanism of the dynamical condensate generation is impossible in the real massive scalar field theory. That should be a non--trivial observation because in this paper we have essentially resummed the leading IR loop corrections, while the Coleman--Weinberg potential follows from the resummation of the leading UV loop contributions.

We would like to acknowledge discussions with A.Alexandrov, K.Bazarov, K.Gubarev, A.Radkevich and A.Semenov. This work was supported by the Foundation for the Advancement of Theoretical Physics and Mathematics “BASIS” grant, by RFBR grants 19-02-00815 and 21-52-52004, and by Russian Ministry of education and science.


\vspace{5mm}


\begin{thebibliography}{1}

\bibitem{AGD} Abrikosov, A.A., Gorkov, L.P. and Dzyaloshinskii, I.Y. (1965) Quantum Field Theoretical Methods in Statistical Physics. Pergamon Press, New York.

\bibitem{LL9} L. D. Landau and E. M. Lifshitz, Vol. 9 (Pergamon Press, Oxford, 1975).

\bibitem{Krotov:2010ma}
D.~Krotov and A.~M.~Polyakov,
Nucl. Phys. B \textbf{849}, 410-432 (2011)
doi:10.1016/j.nuclphysb.2011.03.025
[arXiv:1012.2107 [hep-th]].

\bibitem{Moreau:2018lmz}
G.~Moreau and J.~Serreau,
Phys. Rev. Lett. \textbf{122}, no.1, 011302 (2019)
doi:10.1103/PhysRevLett.122.011302
[arXiv:1808.00338 [hep-th]].

\bibitem{Guilleux:2015pma}
M.~Guilleux and J.~Serreau,
Phys. Rev. D \textbf{92}, no.8, 084010 (2015)
doi:10.1103/PhysRevD.92.084010
[arXiv:1506.06183 [hep-th]].

\bibitem{Akhmedov:2013vka}
E.~T.~Akhmedov,
Int. J. Mod. Phys. D \textbf{23}, 1430001 (2014)
doi:10.1142/S0218271814300018
[arXiv:1309.2557 [hep-th]].

\bibitem{Akhmedov:2019cfd}
E.~T.~Akhmedov, U.~Moschella and F.~K.~Popov,
Phys. Rev. D \textbf{99}, no.8, 086009 (2019)
doi:10.1103/PhysRevD.99.086009
[arXiv:1901.07293 [hep-th]].

\bibitem{Akhmedov:2015xwa}
E.~T.~Akhmedov, H.~Godazgar and F.~K.~Popov,
Phys. Rev. D \textbf{93}, no.2, 024029 (2016)
doi:10.1103/PhysRevD.93.024029
[arXiv:1508.07500 [hep-th]].

\bibitem{Akhmedov:2014hfa}
E.~T.~Akhmedov, N.~Astrakhantsev and F.~K.~Popov,
JHEP \textbf{09}, 071 (2014)
doi:10.1007/JHEP09(2014)071
[arXiv:1405.5285 [hep-th]].

\bibitem{Akhmedov:2014doa}
E.~T.~Akhmedov and F.~K.~Popov,
JHEP \textbf{09}, 085 (2015)
doi:10.1007/JHEP09(2015)085
[arXiv:1412.1554 [hep-th]].

\bibitem{Akhmedov:2020haq}
E.~T.~Akhmedov and O.~Diatlyk,
JHEP \textbf{10}, 027 (2020)
doi:10.1007/JHEP10(2020)027
[arXiv:2004.01544 [hep-th]].

\bibitem{Akhmedov:2021rhq}
E.~T.~Akhmedov,
[arXiv:2105.05039 [gr-qc]].

\bibitem{LL10} L. D. Landau and E. M. Lifshitz, Vol. 10 (Pergamon Press, Oxford, 1975).

\bibitem{Zubarev} D.N.Zubarev, V.G. Morozov and G.Ropke, ``Statistical mechanicsof non–equilibrium processes'',Akademie Verlag (Berlin), 1996.

\bibitem{Mattuck} R.D.Mattuck, ``A guide to Feynman diagrams in the many-body problem'', 2nd Edition(McGraw Hill, Inc., 1976).

\bibitem{Radovskaya:2020lns}
A.~A.~Radovskaya and A.~G.~Semenov,
Eur. Phys. J. C \textbf{81}, no.8, 704 (2021)
doi:10.1140/epjc/s10052-021-09382-4
[arXiv:2003.06395 [hep-ph]].

\bibitem{Kamenev} A. Kamenev, “Many-body theory of non-equilibrium systems,” Cambridge, UK: Univ. Pr. (2011)
[arXiv:cond-mat/041229].

\bibitem{Akhmedov:2011pj}
E.~T.~Akhmedov,
JHEP \textbf{01}, 066 (2012)
doi:10.1007/JHEP01(2012)066
[arXiv:1110.2257 [hep-th]].




\end{thebibliography}
\end{document}